# Model-based Iterative Reconstruction for Flat-Panel Cone-Beam CT with Focal Spot Blur, Detector Blur, and Correlated Noise

Steven Tilley II, Jeffrey H. Siewerdsen, J. Webster Stayman

Department of Biomedical Engineering, Johns Hopkins University, Baltimore, MD 21205

**Abstract:** While model-based reconstruction methods have been successfully applied to flat-panel cone-beam CT (FP-CBCT) systems, typical implementations ignore both spatial correlations in the projection data as well as system blurs due to the detector and focal spot in the x-ray source. In this work, we develop a forward model for flat-panel-based systems that includes blur and noise correlation associated with finite focal spot size and an indirect detector (e.g., scintillator). This forward model is used to develop a staged reconstruction framework where projection data are deconvolved and log-transformed, followed by a generalized least-squares reconstruction that utilizes a non-diagonal statistical weighting to account for the correlation that arises from the acquisition and data processing chain. We investigate the performance of this novel reconstruction approach in both simulated data and in CBCT test-bench data. In comparison to traditional filtered backprojection and model-based methods that ignore noise correlation, the proposed approach yields a superior noise-resolution tradeoff. For example, for a system with 0.34 mm FWHM scintillator blur and 0.70 FWHM focal spot blur, using the a correlated noise model instead of an uncorrelated noise model increased resolution by 42% (with variance matched at $6.9 \times 10^{-8}$ $mm^{-2}$). While this advantage holds across a wide range of systems with differing blur characteristics, the improvements are greatest for systems where source blur is larger than detector blur.

## 1 INTRODUCTION

Flat-panel cone-beam CT (FP-CBCT) is a promising modality for many clinical applications due, in large part, to its adaptable open geometry and capacity for high, isotropic spatial resolution. A broad variety of geometries exist, each driven by application requirements as well as tradeoffs between x-ray scatter (favoring greater object-detector distance) and spatial resolution. In some cases, the extended system geometry increases the influence of focal spot blur. Many FP-CBCT systems would benefit from even greater resolution capabilities. Clinical examples in which improved spatial resolution is required include the detection of microcalcifications in CBCT mammography (Gong *et al* 2004, Lai *et al* 2007) and the characterization of trabecular structure in CBCT extremities imaging (Muhit *et al* 2013). In both of these cases, there are important image features that are just beyond the typical spatial resolution of FB-CBCT systems. While the image quality (including the spatial resolution) of current CBCT systems can be improved through hardware changes (smaller detector pixels, thinner scintillator, smaller x-ray source focal spot, etc.) and the redesign of other system characteristics (increased magnification, increased exposure, etc.), improved reconstruction algorithms can also lead to dramatic improvements in image quality. Moreover, improvements to the data processing pipeline have the potential to alter traditional tradeoffs in the design of new systems.

Much work on improved data processing in FP-CBCT has concentrated on improved system modeling for reconstruction. This includes models for data corrections to account for scatter, beam hardening, and source and detector effects (Sisniega *et al* 2015, Siewerdsen *et al* 2006, Ning *et al* 2004, La Rivière *et al* 2006, Zhang *et al* 2014, Hsieh *et al* 2000, Zhu *et al* 2009) as well as changes to the reconstruction algorithm (Dang *et al* 2015, Evans *et al* 2013, Elbakri and Fessler 2002). So-called model-based iterative reconstruction (MBIR) algorithms have demonstrated higher image quality than traditional analytical approaches like filtered backprojection (FBP) in both multi-detector CT (Thibault *et al* 2007) as well as CBCT (Wang *et al* 2014, Dang *et al* 2015, Wang *et al* 2009, Sun *et al* 2015). Much of this success comes from an accurate statistical model of measurement noise. Typically, statistical approaches model the data-dependent variance of measurements and implicitly or explicitly weigh the relative contributions of data with differing noise levels. Nearly all MBIR methods have made the assumption that the measurements are statistically independent. However, a few counterexamples can be found in the literature in Fourier rebinned PET (Alessio *et al* 2003) and in multi-energy CT reconstruction (Sawatzky *et al* 2014, Liu and Yu 2015, Brown *et al* 2015). While the independence assumption may be appropriate for some multi-detector CT systems, FP-CBCT data can exhibit significant spatial noise correlation due to the detection process (e.g., as part of the indirect x-ray detection and light spread in the scintillator). In this work we discuss, model, and integrate the effect of noise correlation in the reconstruction algorithm.

For high spatial resolution reconstructions, accurate modeling of system blur is also potentially important. System blur modeling has been used extensively in MBIR for nuclear imaging to achieve higher spatial resolution reconstructions (Feng *et al* 2006, Chun *et al* 2013, Tsui *et al* 1987, Yu *et al* 2000). Such methods have also been attempted in multi-detector CT (MDCT); however, it should be noted that in many cases, sophisticated blur modeling does not yield significant improvements. For example, Hofmann et al (Hofmann *et al* 2013) showed that, under typical diagnostic CT conditions (0.5-2.0 mm focal spot size, 1.2 mm detector pixel size), modeling the effects of an extended source focal spot was not beneficial. However, there is a large opportunity for improvement in FP-CBCT systems that often have smaller detector pixel pitches, larger (fixed anode) x-ray focal spots, and more varied geometries than clinical multi-detector CT.

Previous work to improve the accuracy of the system model has accounted for blur in the reconstruction algorithm but without modeling noise correlation (Feng *et al* 2006, Yu *et al* 2000). Other staged reconstruction approaches perform a sinogram restoration step to account for system blur, (La Rivière *et al* 2006) and noise correlation (Zhang *et al* 2014, Wang *et al* 2006). We propose a forward model that includes source and scintillator blur and a reconstruction algorithm that tracks noise correlation through a deblurring step and incorporates this correlation in an iterative optimization algorithm. We evaluate the method by comparing reconstructions using the correlated noise model with results obtained using a model that assumes spatially independent noise. This work expands upon previous work where we have shown correlated noise models are advantageous when dealing with high readout noise and no source blur (Stayman *et al* 2014) and in more recent work in which we introduced a regularized deblurring step (Tilley II *et al* 2014). The latter work considered preliminary simulation studies with three different blur scenarios using a small phantom and demonstrated that systems dominated by detector blur with low readout noise do not benefit from the proposed model, but source blur dominated systems do. In this work, we present a complete derivation and development of the deblurring process and correlated noise model reconstruction algorithm. We evaluate the performance under different noise model assumptions and compare reconstructed image quality in a digital extremities phantom in a variety of system configurations with different combinations of source and detector blur sizes. Additionally, we demonstrate performance advantages using the proposed approach in physical FP-CBCT data of an anthropomorphic wrist phantom from an x-ray test bench for which we have measured system blur associated with both the x-ray source and the detector.

## 2 METHODS

*2.1 System Model for Mean Measurements and Covariance*
In Figure 1 we present an idealized model of a FB-CBCT system. The measurements are modeled as a random vector which has undergone a series of transformations as the signal propagates through the system. We presume that the statistical distribution of quanta at each stage is approximately Gaussian and concentrate only on the first- and second-order statistics of this random vector. In each of four stages of the FB-CBCT system we identify the mean (across the bottom of the figure) and the covariance matrix (top) for this random vector.

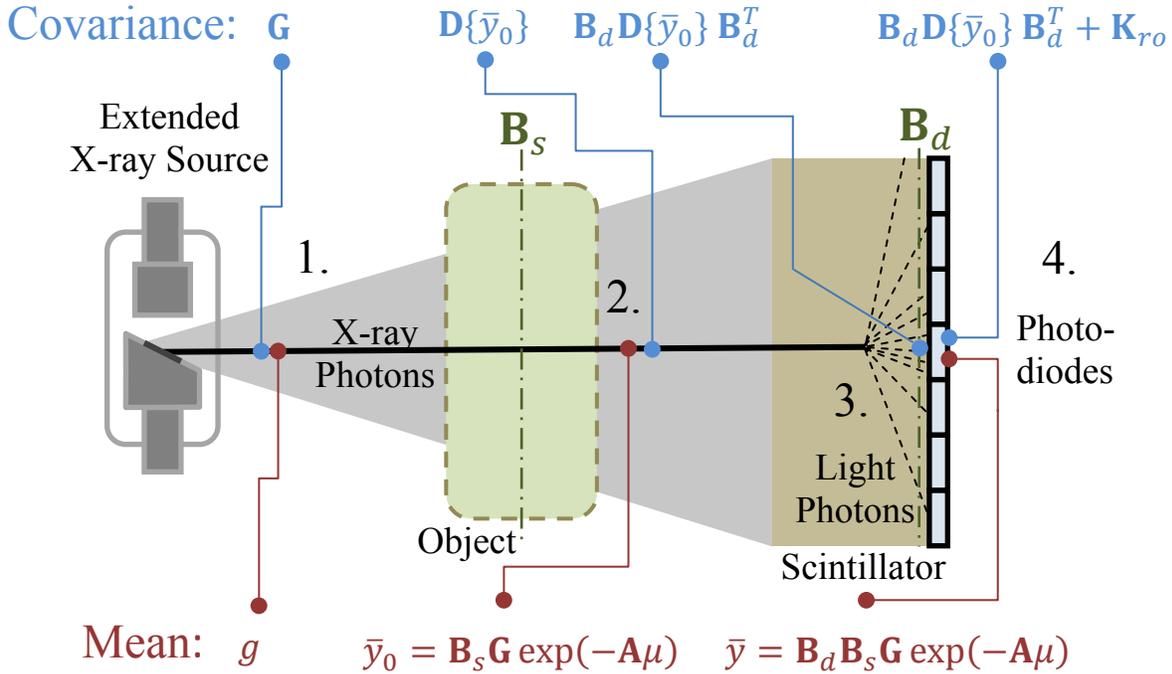

Figure 1: Model for the mean (bottom) and covariance (top) of quanta at various stages. 1) After x-ray photon generation, quanta are independent with a variance equal to the mean. 2) When x-ray photons are attenuated by the object, the spatial distribution of the mean and variance change, but remain equal and independent. An operator that includes source blur is also included. 3) In the scintillator, x-ray photons interact with a scintillating material creating many light photons which spread spatially, blurring the mean distribution and adding correlation to the noise. 4) Photodiodes detect the light photons with possible additive readout noise.

In the first stage, we presume the X-ray tube generates a spatial distribution of x-ray photons with a mean vector $g$. These primary quanta are independent with a Poisson distribution. Thus, the covariance is given by a diagonal matrix with the vector $g$ on the diagonal.

At the second stage, x-ray photons have been attenuated by the object in accordance with Beer's law and the signal has been blurred by the extended x-ray source. The pre-detection mean distribution of x-ray photons is given be the vector $\bar{y}_0$. Similarly, since source blur does not correlate noise (Moy 2000), the covariance matrix remains diagonal with values updated to match the mean vector. (Note that we use the diagonal operator $\mathbf{D}\{\cdot\}$ throughout the paper that puts its vector argument on the main diagonal of a square matrix. For example, $\mathbf{D}\{\bar{y}_0\}$ indicates a diagonal matrix with vector $\bar{y}_0$ along the diagonal). The mean vector $\bar{y}_0$ is modeled using Beer's law along line integrals obtained by applying a forward projection operator ($\mathbf{A}$) to the vector of attenuation values ($\mu$). A scaling by the primary quanta distribution is also applied ($\mathbf{G}$).

Focal spot blur is modeled as a linear operator ($\mathbf{B}_s$) on the (unblurred) transmission values. Ideally, this focal spot blur might be modeled by an integration of transmission values over a fine sampling of rays distributed over the extended focal spot, which would accommodate the depth-dependent blur associated with the source. However, while the above mathematical model is general, a model with very fine sampling is expensive to compute. In this work we presume that the object being scanned has a small width relative to the source-detector distance so that $\mathbf{B}_s$ may be approximated as a single convolutional blur function acting within the object plane (at the source-object distance). This blur is applied on each projection using a system model with the number of rays equal to the number of detector elements. Moreover, we choose a separable footprints projector/backprojector pair (Long *et al* 2010) that explicitly models the detector aperture and cubic voxels. This approximation moves the physical integration over the detector aperture (ideally modeled in $\mathbf{B}_s$ or $\mathbf{B}_d$) inside the exponential (as part of **A**). While not strictly correct, this approximation accounts for a degree of blur due to the detector aperture at reduced computational cost.

In the detector scintillator (stage 3), individual X-ray photons are converted into many visible light photons with a broad angular distribution in trajectories. This results in a single X-ray photon contributing signal to multiple pixels. This is modeled as a second blurring operator ($\mathbf{B}_d$) which modifies the mean vector. Because of the one-to-many conversion of primary to secondary quanta in the detector, the scintillator also correlates the noise associated with each x-ray quanta, resulting in a non-diagonal covariance matrix.[1] In this work, it is assumed that any blur associated with the pixel aperture is negligible compared to scintillator blur and that aliasing is not a dominating effect.

At the final stage (position 4), we include additional zero-mean electronic readout noise associated with the detector. This is modeled by adding the readout noise covariance matrix ($\mathbf{K}_{ro}$) to the current covariance matrix. Thus, the final mean and covariance of the measurement random vector are given by equations (1) and (2).

$$\bar{y} = \mathbf{B}_d \mathbf{B}_s \mathbf{G} \exp(-\mathbf{A}\mu) \tag{1}$$
$$\mathbf{K}_Y = \mathbf{B}_d \mathbf{D}\{\mathbf{B}_s \mathbf{G} \exp(-\mathbf{A}\mu)\}\mathbf{B}_d^T + \mathbf{K}_{ro} \tag{2}$$

For convenience, we will further define total system blur as

$$\mathbf{B} \triangleq \mathbf{B}_d \mathbf{B}_s \tag{3}$$

*2.2  Data Preprocessing and Linearization*

If an estimate, $\hat{l}$, of the true line integrals, *l*, were available, estimating $\mu$ may be framed as a linear problem ($l = \mathbf{A}\mu$). Under a presumption of Gaussian noise, the maximum-likelihood solution may be found according to a generalized weighted least-squares objective, where the inverse of the covariance matrix is the weighting matrix. To obtain such an estimate of the line integrals we can consider the following processing of the raw measurement data:

$$\hat{l} = -\log\left(\mathbf{G}^{-1}\mathbf{C}'y\right) \tag{4}$$

where we introduce a generic deblurring operation, $\mathbf{C}' \approx \mathbf{B}^{-1}$. We note that an exact inversion of **B** may not be desirable or even possible depending on the form of the total system blur. This will be discussed in greater detail below. Naturally, the transformation in (4) changes the covariance structure. One can show that the covariance for the line integrals may be approximated as:

$$\mathbf{K}_L \approx \mathbf{D}\left\{\frac{1}{\mathbf{C}'\bar{y}}\right\}\mathbf{C}'\mathbf{K}_Y[\mathbf{C}']^T\mathbf{D}\left\{\frac{1}{\mathbf{C}'\bar{y}}\right\} \tag{5}$$

This approximate form is derived using a Taylor approximation and is described in detail in the appendix. It is this expression that we will use for noise modeling in a statistical reconstruction method.[2]

One must take care in choosing the exact form for the deblur operator $\mathbf{C}'$. If **B** is not full rank, then there is a null space that cannot be recovered. Similarly, if **B** is highly ill-conditioned with near zero singular values, extreme noise amplification and sensitivity to finite precision computing may occur. To avoid these scenarios, we introduce a modified deblur operator below that avoids these undesirable features.

---

[1] In reality, this blur is stochastic. In this work we assume the randomness associate with the blur is negligible.
[2] While not a focus of this work, the above mean measurement and covariance models could potentially also be used to design improved deblurring methods for 2D radiography.

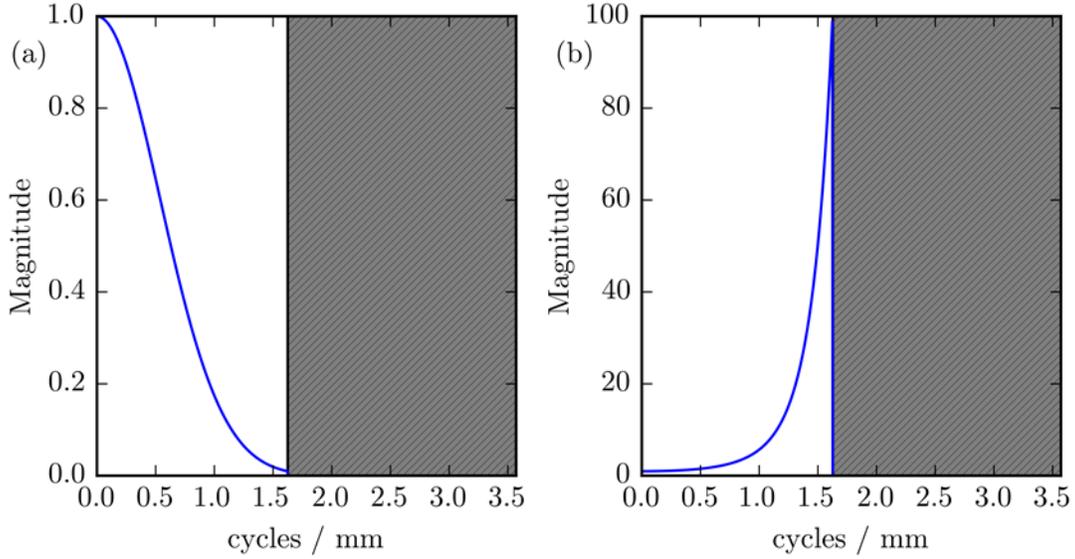

**Figure 2**: Representation of deblur masking. The original blur (a) tends towards zero at higher frequencies. To account for this, those frequencies are masked. Figure 2(a) shows $B^+$ from equation 7 and Figure 2(b) shows $B^-$ from equation 6.

In this work, we assume the system blur is shift-invariant, which allows blurring and deblurring operations to be performed using Fourier methods. In this case, zero or near zero singular values associated with null-spaces and ill-conditioning can be identified at specific spatial frequencies in the transfer function, since Fourier operators diagonalize the circulant system blur operator, **B**. Thus, one simple solution is to mask the nulled or near null frequencies. If $B$ is the unmodified Fourier transfer function such that

$$\mathbf{Q}^*\mathbf{D}\{B\}\mathbf{Q} = \mathbf{B}$$

where **Q** is the discrete Fourier transform, then we may apply a modified deblur operator with threshold parameter $\epsilon$ as

$$\mathbf{C}' \triangleq \mathbf{Q}^*\mathbf{D}\{B^-\}\mathbf{Q} \quad [B^-]_i = \begin{cases} 0 & \text{if } \left|\frac{B_i}{B_0}\right| < \epsilon \\ 1/B_i & \text{otherwise} \end{cases} \tag{6}$$

where $B_i$ are individual frequency components of the blur, and $B_0$ is the zero frequency component (unity for an energy preserving blur). Similarly, we may define an approximate inverse to this thresholded deblur operation as

$$[\mathbf{C}']^{-1} \approx \mathbf{C} \triangleq \mathbf{Q}^*\mathbf{D}\{B^+\}\mathbf{Q} \quad [B^+]_i = \begin{cases} 0 & \text{if } \left|\frac{B_i}{B_0}\right| < \epsilon \\ B_i & \text{otherwise} \end{cases} \tag{7}$$

The masked deblur and blur transfer functions are illustrated in Figure 2. A representation of (7) is shown in Figure 2(a), which is a blur transfer function with the high frequencies masked out, as indicated by the grey hatched area. A representation of (6) is shown in Figure 2(b), which is an inversion of (7) within the unmasked region, and zero otherwise.

The diagonal terms in (5) can easily be inverted by inverting each element on the diagonal. This fact and (7) give expressions for the inverse of each term in (5) except $\mathbf{K}_Y$. The inverse of (5) can therefore be written as:

$$\mathbf{K}_L^{-1} \approx \mathbf{D}\{\mathbf{C}'\bar{y}\}\mathbf{C}^T\mathbf{K}_Y^{-1}\mathbf{C}\mathbf{D}\{\mathbf{C}'\bar{y}\} \tag{8}$$

$\mathbf{K}_Y$ is approximated as:

$$\mathbf{K}_Y = \mathbf{C}_d \mathbf{D}\{\mathbf{C}'y\}\mathbf{C}_d^T + \mathbf{K}_{ro} \tag{9}$$

where $\mathbf{C}_d$ is a thresholded scintillator blur. This thresholded blur is defined in (7), with $\mathbf{B}_d$ replacing $\mathbf{B}$. The diagonal term from equation (2), which requires unavailable mean pre-detection data, is estimated by the deblurred measurement data.

## 2.3 Penalized Generalized Weighted Least-Squares Reconstruction

With the previously described processing in (4) which linearizes the system model, and the presumption of Gaussian distributed noise with zero mean and known covariance, we may form an objective function. Under these assumptions the implicitly defined objective is a generalized least-squares fit with a weighting by the inverse of the covariance matrix. Specifically, we may write

$$\hat{\mu} = \arg\min_{\mu} \|\hat{l} - \mathbf{A}\mu\|^2_{\mathbf{K}_L^{-1}} + \beta R(\mu) \tag{10}$$

where $\hat{\mu}$ is a vector of the estimated attenuation values. We have included a regularization term $R(\mu)$ and regularization strength parameter $\beta$ to control the trade-off between noise and resolution in the reconstruction. In this work, we apply a quadratic penalty on pairwise voxel differences:

$$R(\mu) = \frac{1}{4}\sum_i\sum_j w_{ij}(\mu_j - \mu_i)^2 = \frac{1}{2}\mu^T \mathbf{R}\mu \tag{11}$$

where the weights $w_{ij}$ indicate the relative strength of the penalty for each pixel pair. (We concentrate on a penalty where first-order neighbors have unity weights; otherwise $w_{ij}$ equals zero.) The penalty can also be written in matrix form in (11), where $\mathbf{R}$ is a convolution matrix. Thus, with a quadratic penalty, the solution to (10) can be rewritten explicitly as

$$\hat{\mu} = \arg\min_{\mu} \|\hat{l} - \mathbf{A}\mu\|^2_{\mathbf{K}_L^{-1}} + \frac{1}{2}\beta\mu^T\mathbf{R}\mu = (\mathbf{A}^T\mathbf{K}_L^{-1}\mathbf{A} + \beta\mathbf{R})^{-1}\mathbf{A}^T\mathbf{K}_L^{-1}\hat{l} \tag{12}$$

Sauer and Bouman derived this general form (with a diagonal weighting matrix) using a second order Taylor series approximation to a likelihood function based on a Poisson noise model and a Gaussian Markov random field prior (Sauer and Bouman 1993). Fessler modified the derivation to include scatter, and used a more general penalty term (Fessler 1995).

Beside the additional deblurring step in (4), the key difference between equation (12) and the work of Sauer, Bouman, and Fessler, is the use of a non-diagonal weighting matrix - the inverse of $\mathbf{K}_L$ - which may not necessarily exist. For the formulation used above, the masked frequencies in $\mathbf{C}$ result in a degenerate $\mathbf{K}_L$ which cannot be inverted (i.e. information at certain frequencies cannot be recovered). If we replace the inverse in (10) with a generalized inverse, the unregularized solution to (10) is the maximum likelihood solution in the span of $\mathbf{K}_L$. At line integral frequencies in the null space, the likelihood function evaluates to zero and the reconstruction relies only on the penalty function. In this work, we will use inverse notation to mean generalized inverse where appropriate. See (Rao 1973) for a detailed explanation of Gaussian density functions with degenerate covariance matrices.

Equation (12) can be logically separated into two sections (see Algorithm 1). The preprocessing section requires estimating the line integrals, applying the inverse covariance matrix, and backprojecting (applying $\mathbf{A}^T$ to) the result. The application of the inverse covariance matrix is performed in steps in

Preprocessing Section:
$y_{db} \leftarrow \mathbf{C}'y$ % Apply inverse source and detector blur
$\hat{l} = -\log(\mathbf{G}^{-1}y_{db})$ % Perform normalization and logarithmic transformation
$z = \mathbf{K}_L^{-1}\hat{l}$ % see Evaluating $\mathbf{K}_L^{-1}$
$b = \mathbf{A}^T z$ % Precompute backprojected transformed line integrals

Iterative Section:
Solve $\mathbf{M}\mu = b$ % see Evaluating $\mathbf{M}$.
  Conjugate Gradient % $N_\mu$ iterations (outer loop)

Evaluating $\mathbf{K}_L^{-1}x$ (subroutine):
$p = [\mathbf{C}]^T \mathbf{D}\{y_{db}\} x$
Solve $\mathbf{K}_Y q = p$
  Conjugate Gradient % $N_k$ iterations (inner loop)
$b = \mathbf{D}\{y_{db}\}[\mathbf{C}] q$
  Return $b$

Evaluating $\mathbf{M}x$ (subroutine):
  Return $\mathbf{A}^T \mathbf{K}_L^{-1} \mathbf{A}x + \beta \mathbf{R}x$

**Algorithm 1: Overview and pseudocode for proposed algorithm.**

accordance with (8). The iterative section involves two iterative optimization algorithms, the "outer" system algorithm, which solves for $\hat{\mu}$, and the "inner" inversion of $\mathbf{K}_Y$ within $\mathbf{K}_L^{-1}$. The inversions of $\mathbf{K}_Y$ and $(\mathbf{A}^T \mathbf{K}_L^{-1} \mathbf{A} + \beta \mathbf{R})$ can be performed using a variety of solvers, such as LSQR($\mathbf{A}^{-1}$), (Benbow 1999) or a momentum based approach (Nesterov 1983, 2005, Kim *et al* 2015). In this work both systems were solved using the conjugate gradient method (Hestenes and Stiefel 1952).

The actions of $\mathbf{A}$ and $\mathbf{A}^T$ were performed on a GPU using the separable footprints algorithm (Long *et al* 2010) and an in-house CUDA library (Nickolls *et al* 2008). The rest of the code was written in python (python.org) using scipy, (Jones *et al* 2001, Walt *et al* 2011) and plots were generated using Matplotlib (Hunter 2007).

To evaluate our new approach, we will compare against a more traditional (uncorrelated) noise model and FBP. The uncorrelated noise model approximates an independent Poisson model, given by the diagonal covariance matrix:

$$\mathbf{K}_L^{uncorr} = \mathbf{D}\{\frac{1}{\mathbf{C}'y + \sigma_{ro}^2}\} \quad (13)$$

Additionally, for comparison with traditional FBP, reconstructions were performed on both deblurred and non-deblurred data using the FDK algorithm (Feldkamp *et al* 1984) using an unapodized ramp filter with a cutoff at the Nyquist frequency and no additional apodization.

*2.4 Simulation Studies*

To investigate the performance of the proposed reconstruction framework, simulation studies were conducted using the digital phantom illustrated in Figure 3(a). This phantom contains a number of different regions that include the following tissue types and attenuation values: (i) fat ($\mu = 0.01875$ mm$^{-1}$); (ii) muscle ($\mu = 0.02150$ mm$^{-1}$); and (iii) bone ($\mu = 0.06044$ mm$^{-1}$). There are additional features for qualitative and quantitative performance analysis, specifically a medium-contrast (v) disc and (iv) line pairs ($\mu = 0.03$ mm$^{-1}$). Simulated mean projection data were generated for a C-arm geometry (120 cm source detector distance and 60 cm source axis distance) with a one-dimensional detector with 1750 pixels and a 0.14 mm pixel pitch. To approximate a continuous domain projection operator, line integrals were obtained by projecting a high resolution version of the phantom (4000x4000 with 0.025 mm voxels) onto a high resolution detector (7000 pixels with 0.035 mm pixel pitch) followed by an integration (over four detector sub-elements) to match the detector grid.

Noisy projection measurements were generated using Gaussian noise. The readout noise ($\sigma_{ro}$) was 1.9 photons and the gain was a constant $10^6$ photons per detector element. Noiseless projection data were blurred by applying $\mathbf{B}_s$ to obtain an intermediate mean vector $\bar{y}_0$. This vector was then blurred by applying $\mathbf{B}_d$ to obtain the noiseless measurement data. Quantum noise was modeled as zero mean Gaussian noise with variance equal to $\bar{y}_0$. This noise was blurred by $\mathbf{B}_d$ to add correlations and then added to the mean vector. Lastly, zero mean Gaussian readout noise with a variance of $\sigma_{ro}^2$ was added to yield the noisy measurement data.

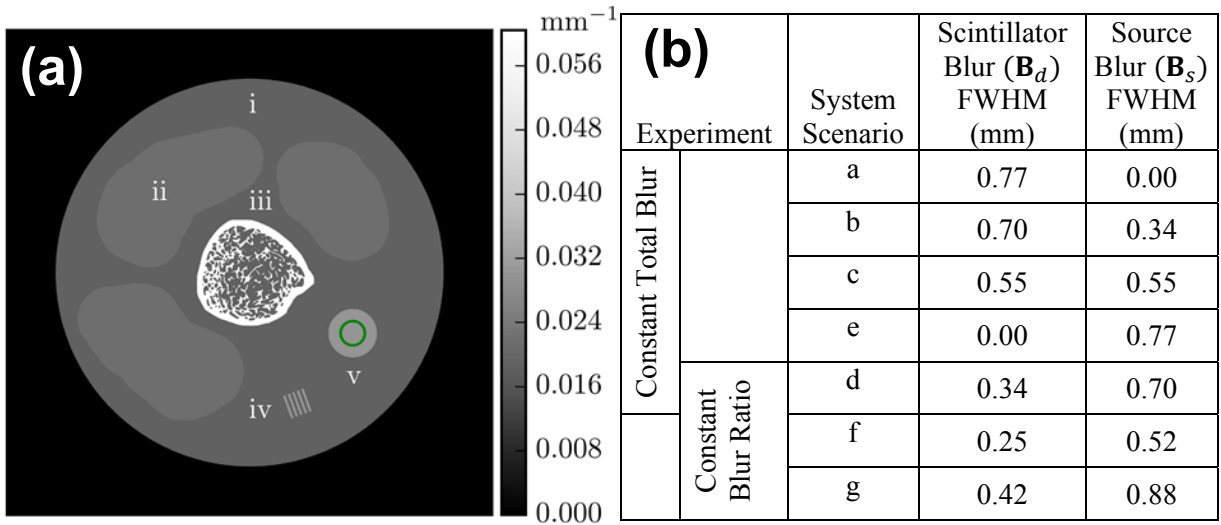

Figure 3: (a) Digital phantom used in simulation studies emulating an extremity imaging scan with i) fat, ii) muscle, iii) bone, iv) line pairs, and v) a uniform disc. For noise evaluations, sample variance was calculated in the disc interior indicated with a circle. Spatial resolution was estimated using the edge response between the disc and fat background. (b) Systems with varying degrees of source and detector blur were simulated to investigate reconstruction performance over a range of scenarios. These scenarios are lettered a-g and permit two experiments where 1) the total blur is constant and the proportion of source and detector blur is varied; and 2) the proportion of source and detector blur is constant and the total blur is varied.

| Experiment | System Scenario | Scintillator Blur ($\mathbf{B}_d$) FWHM (mm) | Source Blur ($\mathbf{B}_s$) FWHM (mm) |
|---|---|---|---|
| Constant Total Blur | a | 0.77 | 0.00 |
| | b | 0.70 | 0.34 |
| | c | 0.55 | 0.55 |
| | e | 0.00 | 0.77 |
| Constant Blur Ratio | d | 0.34 | 0.70 |
| | f | 0.25 | 0.52 |
| | g | 0.42 | 0.88 |

Data were reconstructed into a 1000x1000 two-dimensional image volume with 0.1 mm voxels for each of the three reconstruction methods: PWLS with the proposed correlated noise model, PWLS with the uncorrelated noise model, and FBP. In the preprocessing section $N_k$ was 1000 iterations, and in the iterative section $N_k$ was 100 iterations. $N_\mu$ was 100 iterations. The CG method was terminated early if the residual vector reached zero. Before deblurring, the data were padded with $I_0$. Both $\mathbf{C}_d$, $\mathbf{C}$, and $\mathbf{C}'$ had a threshold ($\epsilon$) equal to $10^{-2}$. When performing covariance operations, padding prior to blur operations was performed using nearest neighbor extrapolation.

For performance assessment in simulation studies, noise-resolution tradeoffs were investigated for each reconstruction approach. Specifically, resolution-variance curves were obtained by sweeping the regularization parameter $\beta$ across a range of values (e.g. lower $\beta$ induces a higher resolution image with more noise, higher $\beta$ yields an image with lower resolution and less noise). To quantify resolution, the width of the edge response of the disc (Figure 3(a)) was estimated using an error function fit. Specifically, attenuation values, $\mu_j$, from a noiseless reconstruction were fit to the following equation which is a function of distance, $x_j$, from the center of the disc in the phantom (from 0.1 mm to 10.0 mm):

$$\mu(x_j) = a + b \operatorname{erf}\left(\frac{x_j - d}{4\sqrt{\log(2)}\, FWHM}\right) \tag{14}$$

The full width at half maximum (FWHM) is derived from this fitting operation. Noise was quantified as the sample variance of attenuation values inside the disc (within a 2.5 mm radius indicated by the ring in Figure 3) for a noisy data reconstruction.

To assess how performance varies with different blur properties, data generation and reconstruction were performed with Gaussian source and scintillator blurs of various sizes (Figure 3(b)). Seven scenarios were chosen, five in which the total system blur was constant and three in which the ratio of the blurs was constant (with one scenario belonging to both groups). Evaluation of systems with constant total blur and varying levels of source and detector blur permits an investigation into the relative performance of methods under varying levels of noise correlation. (Recall that source blur does not introduce noise correlation, whereas detector blur does impart correlation). The constant blur ratio scenarios permit investigation of performance when the total amount of blur is varied. In all studies, for all deblurring and reconstructions, blur models were matched with those used in data generation.

The original 4000x4000 voxel phantom was downsampled by a factor of 4 in both dimensions to obtain a truth image with the same dimensions as the reconstructions. This truth image was used to generate difference images to visualize the accuracy of different reconstructions.

## 2.5 Bench characterization

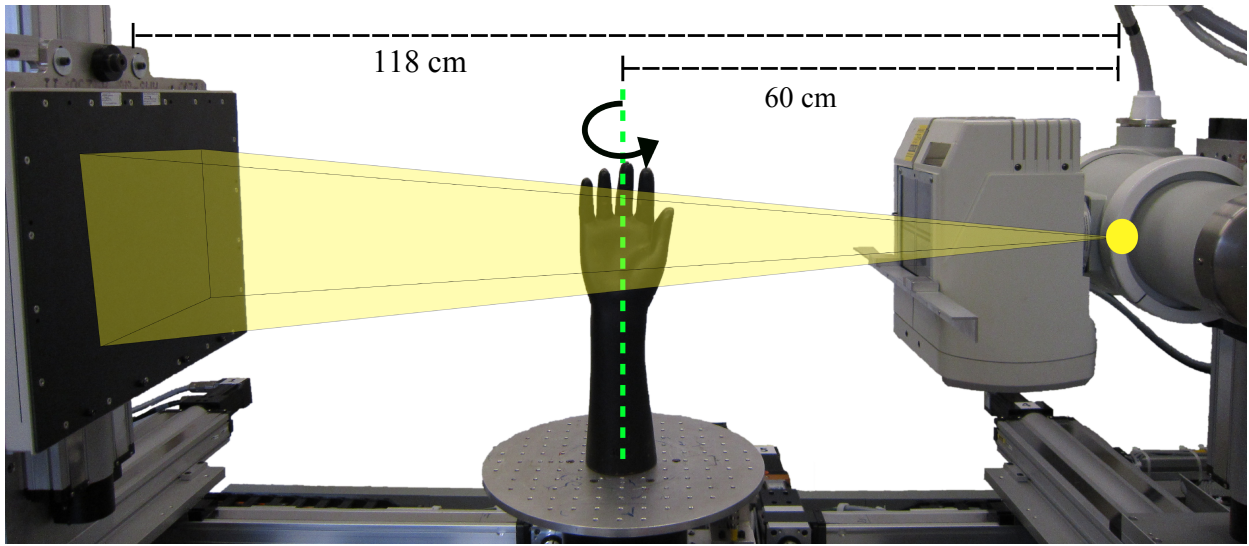

**Figure 4: Test bench with flat panel detector (left) and x-ray source (right). The wrist phantom used is shown at the axis of rotation.**

To apply the proposed methodology to physical data, a system characterization was necessary to estimate system blurs. The experimental setup and CBCT test-bench used for investigations is illustrated in Figure 4. This system is composed of a flat-panel detector (4030CB, Varian, Palo Alto CA) and an x-ray tube (Rad-94, Varian, Salt Lake City UT). Modulation transfer function (MTF) measurements were acquired using a technique similar to that of (Samei *et al* 1998). Multiple projections (720) of a tungsten edge were acquired and gain/offset corrected. These projections were then averaged to reduce noise. The location of the edge was found by fitting error functions to the pixel values along either rows or columns, and fitting a line to the center points of the error functions. The location of this edge was then used to extract the edge-spread function (ESF), which was then binned, differentiated, and Fourier transformed to obtain the MTF.

Both detector and source MTF were computed. The detector MTF was acquired by placing the tungsten edge on the face of the flat-panel detector. The MTF was modeled as a Gaussian (scintillator) multiplied by a sinc. This model was fit to the measured MTF by varying the width of the Gaussian. It was assumed that the scintillator blur is radially symmetric, so only one edge orientation was needed. The source MTF was acquired by placing the tungsten edge at isocenter, and rotating the edge about the source-detector axis to measure different slices of the MTF. The edge was (approximately) oriented to obtain an edge response along the axis of rotation (axial), perpendicular to the axis of rotation (trans-axial), and at 45° to the axis of rotation. These experiments yielded the composition of both source and detector blur. Thus, to find the source blur, the combined source and detector MTF was divided by the detector MTF. Gaussians were fit to the main lobe of the axial and trans-axial slices of the source MTF, and combined into a separable 2D MTF estimate in the reconstruction algorithm.

As an additional check on the nature of the source blur, focal spot images were obtained. A pinhole image of the focal spot was taken using a large source-detector distance to obtain a source magnification of approximately 14. Multiple projections (100) were acquired and averaged. The background of the pinhole image was de-trended by fitting a paraboloid to the background and subtracting it from the entire image. Fourier transforming the pinhole image permitted profiles of the 2D MTF to be compared with the measurements obtained using the tungsten edge.

*2.6    Test-Bench Data Reconstructions*

To investigate the performance of the proposed reconstruction algorithm on physical data, we scanned a custom wrist phantom (The Phantom Laboratory, Greenwich, NY). This phantom includes a natural human skeleton of the arm, wrist, and hand bones in a tissue-equivalent plastic including simulated cartilage and tendon features. Projections were obtained over 720 angles in a 360° circular orbit in a C-arm geometry (source to detector distance of 118 cm and source to axis distance of 60 cm). The reconstruction volume was 600 x 600 x 210 voxels -with cubic voxels 0.15 mm to a side. The x-ray tube on the test-bench had two focal spot settings. All projection data for system characterization and reconstruction comparisons were acquired with the large focal spot (0.8 specification). However, one acquisition was also obtained using the small focal spot (0.4 specification). These data were reconstructed using filtered backprojection to generate a high-resolution reference image with which to compare images from the various reconstructions of projection data with larger focal spot blurs. Data were preprocessed according to the methodology in Section 2.2; however, a few additional calibrations were required for the physical data. Specifically, following traditional gain and offset correction and individual frame normalization of the projection data, the model gain term associated with the primary quanta was estimated. The gain of the system was estimated as a constant equal to the ratio of the mean and variance of an air portion of the normalized projection data (i.e. fitting the Poisson assumption). Data were padded with nearest neighbor values prior to deblurring. Subsequent preprocessing of the data was as described in Section 2.2, with the addition of a thresholding operation on the deblurred data so the minimum was approximately equal to the measurement that would be expected for x-rays attenuated by 40 cm of water.

The data were reconstructed as previously described, using FDK, the uncorrelated noise model, and the correlated noise model. Gaussian approximations for source and scintillator blur were used for deblurring and applying $\mathbf{K}_L^{-1}$. For the proposed reconstruction with a correlated noise model, $\mathbf{C}_d$ had a threshold ($\epsilon$) of zero and padding with zeros. Readout noise was estimated as 1.9 photons. The threshold ($\epsilon$) for $\mathbf{C}$ and $\mathbf{C}'$ was $10^{-2}$ with data padded using nearest neighbor extrapolation. $N_k$ was 1000 iterations in the preprocessing section and 100 iterations in the iterative section. $N_\mu$ was 300 iterations.

Performance evaluation of the physical data reconstructions was conducted using qualitative comparisons to each other and the high resolution reference. Spatial variance was measured in a constant region of a center slice, and used to noise match the reconstructions (by choosing appropriate values for $\beta$) obtained using iterative methods for fair comparison.

# 3 RESULTS

## 3.1 Simulation Studies

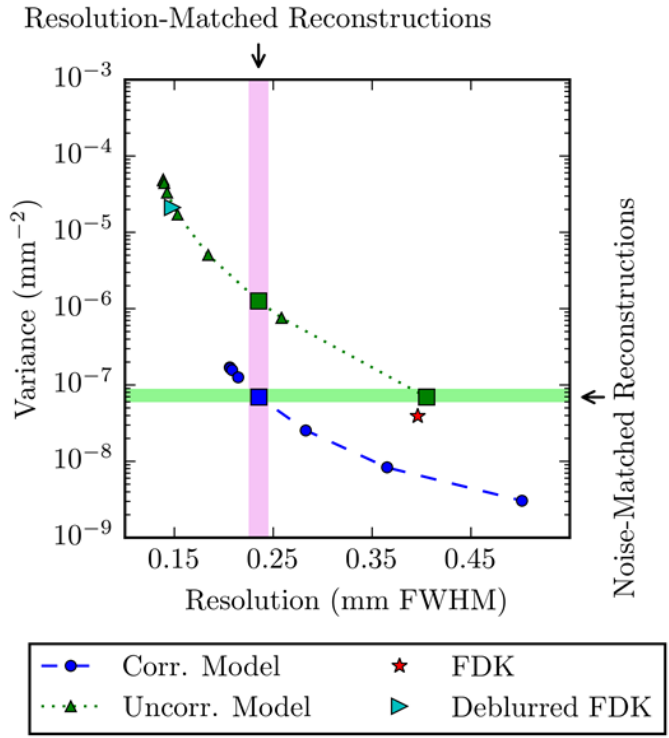

Figure 5: Spatial resolution-variance tradeoff for different reconstructions in a simulated tomographic system with a 0.34 mm FWHM scintillator blur and a 0.70 mm FWHM source blur. Variance and spatial resolution are shown for PWLS reconstructions using the correlated and uncorrelated noise models and FDK. The square data points in the uncorrelated model dataset are either noise-matched or resolution-matched to the square data point in the correlated model dataset.

Figure 5 shows the spatial resolution-variance tradeoff for PWLS reconstructions using the correlated noise model and the uncorrelated noise model in a simulation study with a 0.34 mm FWHM scintillator blur and a 0.70 mm FWHM source blur (system scenario d). Different positions on each curve were obtained by varying regularization strength ($\beta$). When $\beta$ for each approach is chosen such that the uncorrelated and correlated reconstructions have the same resolution (i.e., the vertically aligned squares in Figure 5 are resolution matched), the uncorrelated model yields a variance more than an order of magnitude larger than that of the correlated noise model. Similarly, when $\beta$ is chosen such that the variance is the same (i.e., horizontally aligned squares in Figure 5 are noise matched), the correlated noise model reconstruction has a smaller edge response than the uncorrelated noise model reconstruction, by about 0.17 mm (i.e., 42% decreased FWHM). Non-deblurred FDK is represented by a star demonstrating the traditional spatial resolution limit when no system blur models are adopted. Both the FDK and deblurred FDK resolution-variance points are close to the uncorrelated model resolution-variance tradeoff.

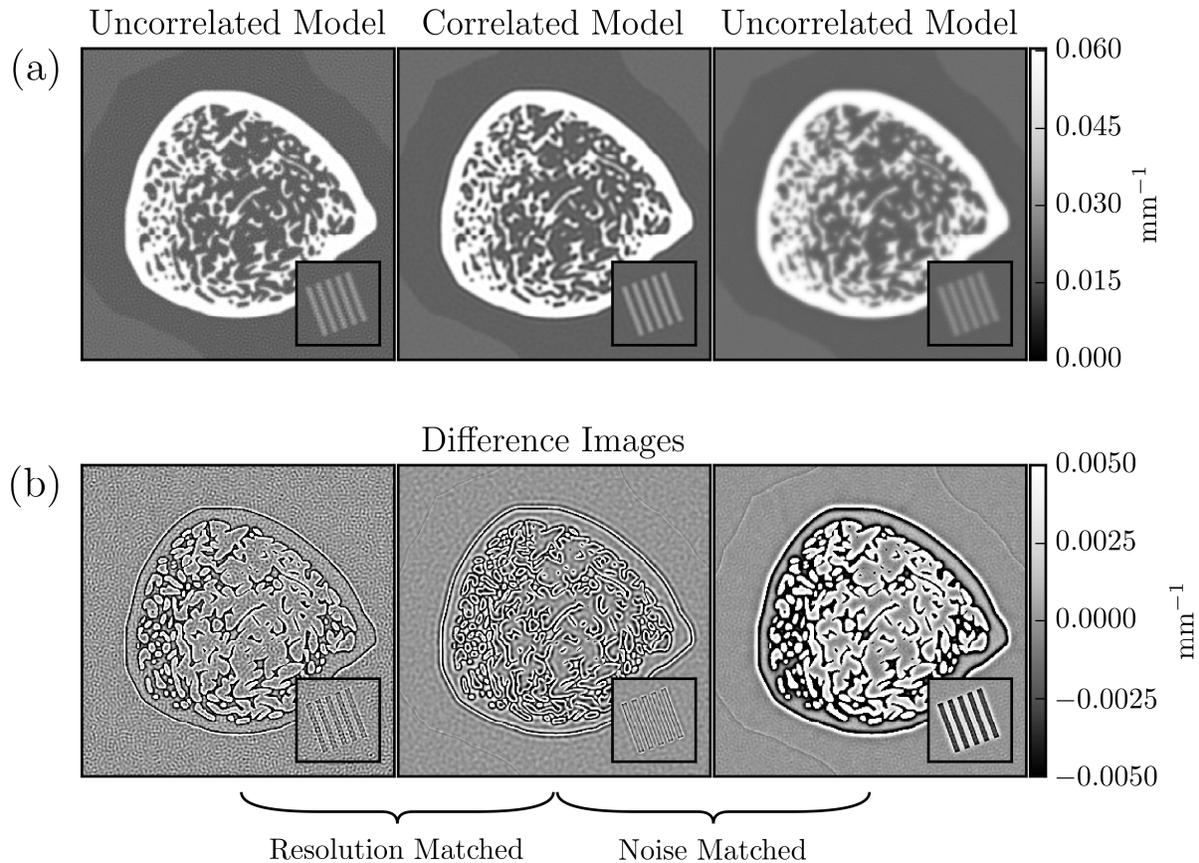

**Figure 6:** (a) Reconstructions from the simulation study corresponding to the noise-matched and resolution-matched data points in Figure 5. (b) Difference images show the difference between the reconstruction and truth. Zoomed images focus on the bone in the center of the phantom, with the line pairs inset in the lower right.

Noise- and spatial resolution-matched reconstructions from Figure 5 are shown in Figure 6(a). These zoomed images show the central portion of the digital phantom with trabecular bone details, as well as an additional zoom inset that shows the line pair object. When the reconstructions are noise-matched, both trabecular details and the line pairs in the uncorrelated reconstruction are blurrier than those in the correlated reconstruction. To better visualize noise, Figure 6(b) shows the difference images where the true image has been subtracted from each reconstruction. In spatial resolution-matched reconstructions, the noise magnitude of the uncorrelated noise reconstruction is larger than that of the correlated noise reconstruction. Note that the noise texture and specific resolution properties also differ between approaches. For example, the correlated noise model produces lower frequency noise than the uncorrelated noise model. Thus, despite matching noise in terms of variance, the noise power spectra clearly differ between the two approaches. Similar observations can be made for spatial resolution. While the FWHM edge response is matched, side lobe performance is clearly different between methods. This is particularly evident in the difference images in the trabecular bone and the line pairs.

To get a better understanding of the performance for different levels of source and detector blur, resolution-variance tradeoffs were evaluated for the seven system blur scenarios delineated in Figure 3(b), represented graphically by the large dots labeled with letters in the top left of Figures 7 and 8. Scenarios a-e (Figure 7) represent systems with constant total blur and varying blur distribution. Specifically, system a is dominated by scintillator blur and system e is dominated by source blur. Scenarios f, d, and g (Figure 8) represent systems in which the ratio of the source and detector blur is constant, but the total blur changes, with f having the smallest blur and g the largest. The remaining plots in Figures 7 and 8 show the resolution-variance curves for each scenario a-e and f, d, and g, respectively. The FDK points indicate the approximate resolution limit without deblurring. Better imaging performance is found toward the bottom and left of each plot (e.g, lower variance and/or higher resolution).

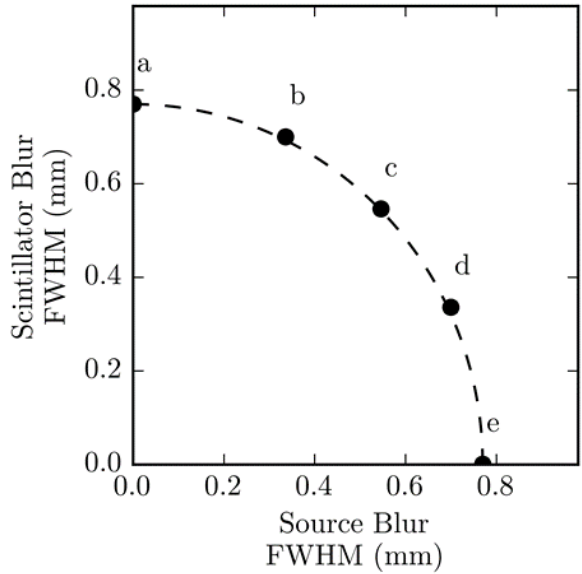

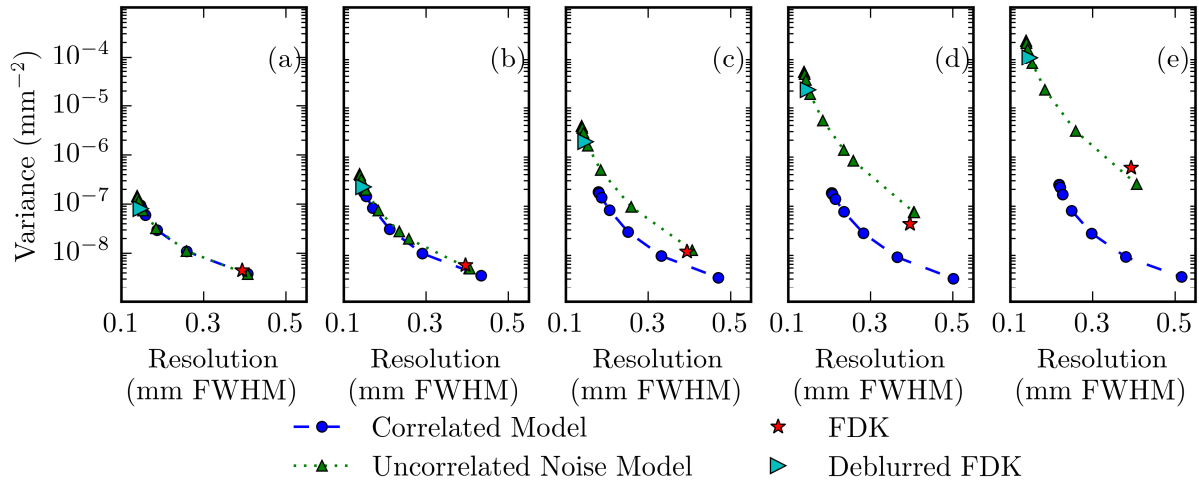

Figure 7: Evaluation of correlated and uncorrelated PWLS reconstruction algorithms for systems with differing source and detector blurs. The schematic on the left illustrates the different imaging system scenarios with varying amounts of source and detector blur. These scenarios have a constant total system blur and a varying blur distribution – from detector-dominated (point a) to source-dominated (point e). The subfigures (a-e) show the resolution-variance tradeoff for each of the systems with blur scenarios corresponding to points a-e in the schematic. For reference, FDK reconstructions are also included in the plots. Note that the correlated noise model shows the greatest advantage for the scenario where source blur dominates.

In all cases, deblurring methods lead to an increase in achievable resolution when compared with non-deblurred FDK. In terms of the resolution-variance tradeoff, the correlated noise model is equivalent to or better than the uncorrelated noise model in all seven cases. When total blur is constant (a-e), the correlated noise model yields the greatest advantage when the blur is due to the source, and yields essentially no advantage when the blur is due only to the scintillator. Intuitively, when source blur is negligible, the deblurring operation is only removing the scintillator blur, thereby whitening the data. In this case, a diagonal covariance matrix becomes an accurate assumption, and the correlated and uncorrelated noise models are essentially equivalent. However, this assumption only holds for low readout noise. As shown in (Stayman *et al* 2014), a system dominated by detector blur will still benefit from a correlated noise model since deblurring will add correlations due to additive readout noise. The FDK and deblurred FDK resolution-variance points lie near the uncorrelated model's resolution-variance curve indicating similar performance. While one often sees an advantage of statistical methods over direct approaches like FDK, the similar performance in this case suggests that the quantum noise modeling[3] does not play a large role in this example (i.e., the dynamic range in variance isn't particularly large for this object), and it is the correlated noise modeling that is important.

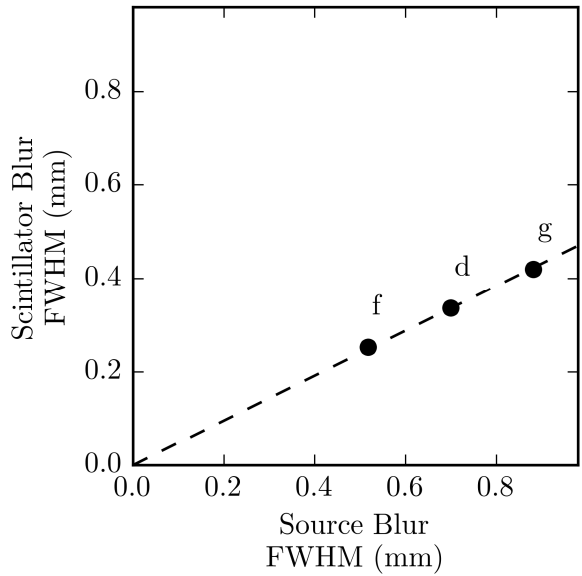
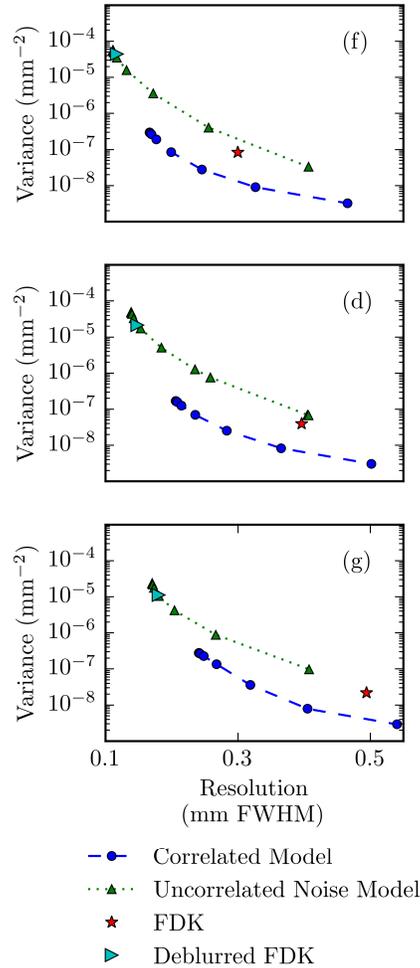

**Figure 8:** Performance evaluation of correlated and uncorrelated models for PWLS reconstructions under varying total blur conditions. The points in the top figure show three different system scenarios, each with the same ratio of source to scintillator blur. The different total system blur conditions represented by points f, d, and g correspond to the resolution-variance plots on the right.

When the system has non-negligible source blur, the deblurring of source blur (which, again, in itself does not correlate the noise) adds correlations to the data, which are accounted for in the correlated noise model but not the uncorrelated noise model, resulting in the advantages seen in Figure 7(c-e). Holding the blur ratio constant while increasing total blur does not have a large effect on the relative performance of the two methods, but does decrease the finest achievable resolution for all methods (Figure 8(f, d, g)).

---

[3] We note that one of the other advantages of model-based approaches is the use of more sophisticated regularization strategies (e.g. non-quadratic penalties). In this work we have concentrated on quadratic penalties and the improvements due to system modeling. Thus, the similar performance of FDK and PWLS with a traditional noise model and quadratic penalty is not completely unexpected.

## 3.2 Bench characterization

Figure 9(a) shows the test-bench detector MTF and the corresponding model approximation. We expect the MTF to be composed of a scintillator MTF and a sinc function (due to pixel sampling). The pixel pitch is 0.388 mm, which would put the first zero of the sinc function at about 2.58 cycles/mm. This is consistent with the measured MTF, which has a first minimum at slightly less than 2.5 cycles/mm. The approximation is a Gaussian multiplied by a sinc of the expected width. The width of the Gaussian in the approximation was found by fitting to the data in the primary lobe. The Gaussian-sinc model is a good approximation to the data at frequencies in this primary lobe.

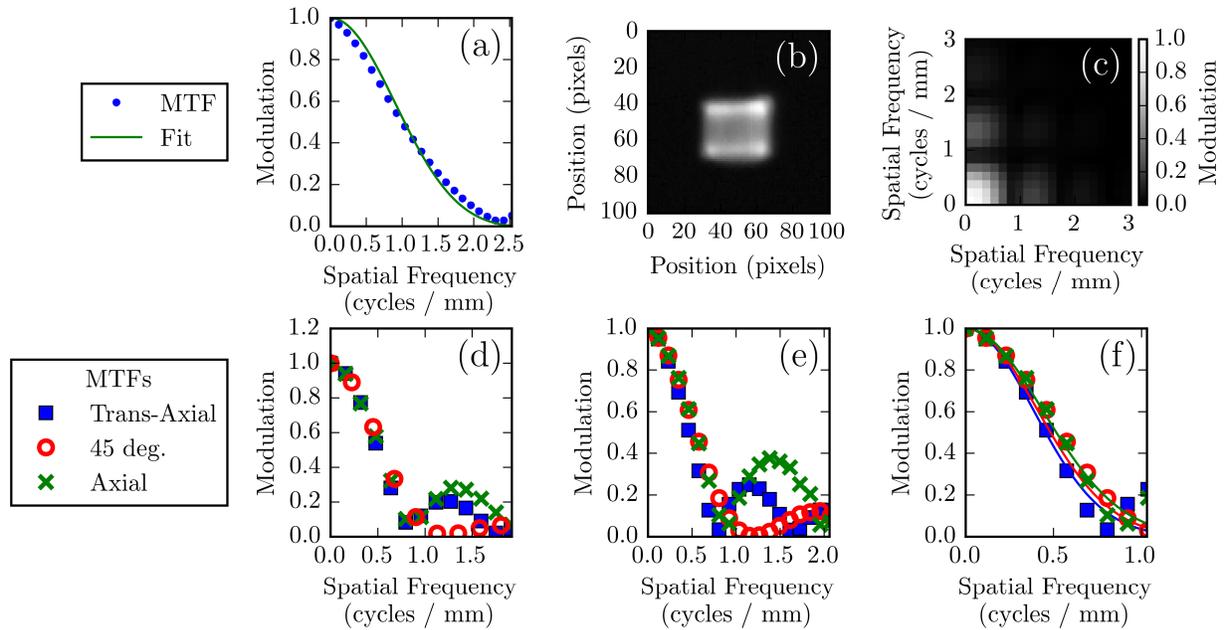

Figure 9: (a) Detector MTF measurements and parameterized fit. The model used for fitting was a Gaussian multiplied by a sinc function. (b) Focal spot image from the CBCT test-bench. (c) A source MTF derived from the focal spot image approximately scaled for focal spot blur at the center of rotation. (d) Trans-axial, axial, and 45° profiles of the pinhole-derived source MTF. (e) Source MTFs as estimated from edge responses at the center of rotation. f) Zoomed version of (e), with solid lines indicating the corresponding profiles of the parameterized fit.

The pinhole image of the source is shown in Figure 9(b). It is approximately a 2D rect function, but a higher order approximation could model the bright horizontal "horns" at the top and bottom of the rect. The pinhole image was Fourier transformed to obtain the 2D MTF shown in Figure 9(c). Along the axial and trans-axial directions the modulation intensity appears similar to a sinc function, consistent with the 2D rect model. The bright horizontal horns at the top and bottom introduce the asymmetry between these two profiles. The source MTF is not radially symmetric, with a noticeably different profile along the 45 degree line. These observations are confirmed in the profiles shown in Figure 9(d), in which modulation intensity is plotted as a function of position along the trans-axial, axial, or 45 degree lines in the 2D MTF.

Figures 9(e-f) show the line profiles of the source MTF acquired using a tungsten edge. The shapes of the MTF profiles are approximately equal to those in Figure 9(d). The axial and trans-axial MTFs are similar to each other at frequencies below the first zero, and the axial MTF is larger in the first side lobe, consistent with the profiles in Figure 9(d). The first zero appears to be at about 0.8 cycles/mm, which corresponds to a 1.3 mm rect function. The 0.8 specification of this spot indicates a focal spot size of 0.8 to 1.1 mm, which is close to our estimate (Bushberg *et al* 2012). The 45 degree MTF has its first zero at a higher frequency, which is consistent with a 2D rect function approximation of the focal spot. This approximation would cause the 45 degree profile to be a squared sinc, resulting in a different shape than the axial and trans-axial profiles, as is seen. The solid lines in 10(e) show the Gaussian approximation used for the test-bench data reconstructions. This model captures most of the shape before the first zero, while ignoring the higher frequencies.

Differences between the MTF derived from the pinhole and edge measurements have two main causes. First, because the exact magnification used to acquire the pinhole image was not known, the frequency axes in 9(c) and 9(d) were scaled so the zeros approximately matched those in Figure 9(e). The MTF derived from the tungsten edge measurement was taken with no source magnification (edge placed at isocenter), so the frequency axis in 9(e) should be considered more accurate. Second, the pinhole image and its derived MTFs were not corrected for detector blur nor for blur associated with the pinhole itself (whose diameter was inexactly known).

### 3.3  *Bench data reconstructions*

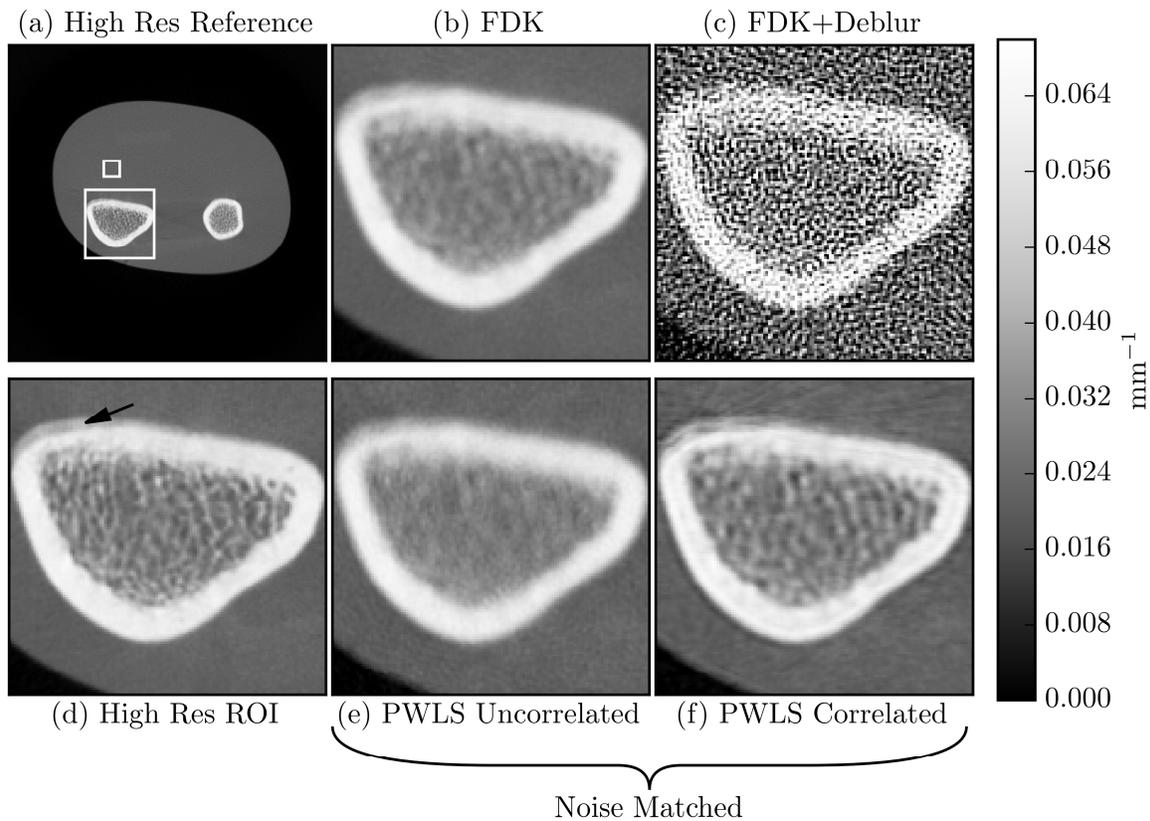

**Figure 10: Test-bench data reconstructions. (a) High resolution reference image. The large box denotes the ROI used for (b-f). (b) FDK reconstruction. (c) FDK reconstruction on deblurred data. (d) High resolution reference image with an arrow indicating cartilage-equivalent plastic. (e) Reconstruction obtained using the uncorrelated noise model. (f) Reconstruction obtained using the correlated noise model and noise matched with (e).**

The correlated noise model was tested on a wrist phantom imaged using the test-bench. Two regions of interest (ROIs) are marked in the high-resolution reference reconstruction (Figure 10(a)) with white rectangles. Figure 10(d) shows the larger ROI (distal radius) from the same reconstruction. This ROI contains cortical bone surrounding trabecular bone. Within this ROI the phantom also contains cartilage-equivalent plastic on the upper left aspect of the cortical bone (marked by an arrow). Images were reconstructed using the FDK algorithm (Figure 10(b)), a combination of deblurring and FDK (Figure 10(c)), the uncorrelated noise model (Figure 10(e)), and the correlated noise model (Figure 10(f)). Figures 10(e) and 10(f) are noise-matched in the smaller ROI indicated in Figure 10(a), with a variance of 8.75 x $10^{-7}$ mm$^{-2}$. The trabecular structure in the FDK reconstruction is present but details are difficult to discern. Deblurring prior to FDK reconstruction results in an unacceptably noisy image. Using the uncorrelated noise model (which includes deblurring) is slightly worse than FDK in terms of resolution at the chosen noise level, although these images are not strictly noise matched. Noting that the uncorrelated PWLS image is both lower resolution and higher noise than FDK underscores the high degree of noise magnification due to the deblurring step. In contrast, the proposed reconstruction method including a correlated noise model recovers more trabecular bone details compared to the noise-matched conventional model-based reconstruction method. We note that the proposed method appears to contain more noise streaks on the upper aspect of the cortical bone. While the exact cause of this increased streaking is unclear and needs additional investigation in future studies, this may be the result of incomplete physical modeling. Specifically, we note that the data was not corrected for beam hardening nor scatter effects and it is possible that blur modeling (and deconvolution) will exaggerate streaking due to these uncompensated biases.

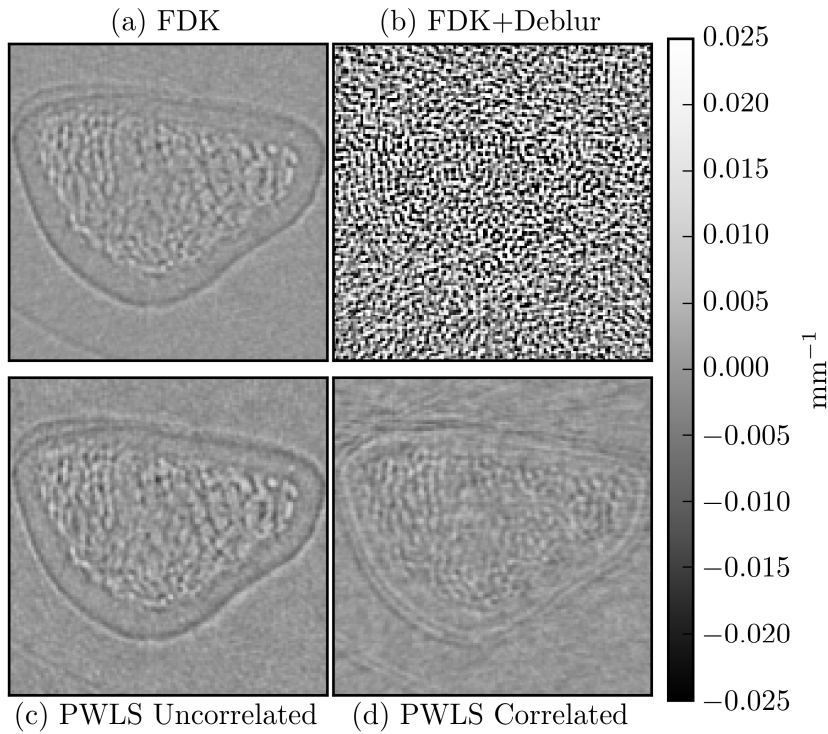

**Figure 11: Difference images between the high-resolution reference image and the various reconstruction methods: (a) FDK; (b) Deblur + FDK; (c) uncorrelated PWLS; and (d) correlated PWLS (corresponding to (b, c, e, and f) in Figure 10). The smallest differences from the high-resolution reference are observed in the correlated PWLS reconstruction.**

Figure 11 shows the difference between the reconstructions in Figure 10(b,c,e,f) and the high-resolution reference reconstruction (Figure 10(d)). The correlated noise model reconstruction is the closest to the high-resolution reference with the flattest difference image. The (non-deblurred) FDK and uncorrelated noise model difference images show quite a bit of structure, indicating a resolution mismatch with the reference image. The FDK reconstruction of deblurred data is overwhelmed by noise. The difference images also show a difference in noise texture between the uncorrelated and correlated model reconstructions, similar to that seen in simulation.

## 4   DISCUSSION

In this work, we have presented an FP-CBCT forward model that accounts for source and detector blurs as well as noise correlation in the data. We used this forward model to develop an MBIR algorithm that uses a staged data processing chain where projections are first deblurred and log-transformed, data correlations are modeled through both the detection and deblurring processes, and reconstruction is performed through a penalized generalized linear least-squares algorithm with a non-diagonal weighting matrix. We have demonstrated the relative performance of the proposed method in comparison to a traditional MBIR approach without a correlated noise model and a filtered backprojection approach with and without deblurring. Experiments in both simulated projections and in CBCT test-bench data demonstrate improved performance of the proposed approach over other methods. Specifically, while any approach that applies deblurring to the projection data permits higher spatial resolution reconstructions, our method can yield significant improvements in noise performance since it maintains an accurate noise model. These improvements vary with the degree of blur and the dominant source of system blur with the greatest advantages for systems with larger focal spot blur.

Despite the advantages illustrated in this work, there are a number of additional opportunities for future work in the development of reconstruction methods that accommodate system blur and data correlation. For example, one could adopt a similar staged deblur and reconstruction method that deconvolves only the correlating (scintillator) blur, whitening the data, and thereby permitting an objective function that accounts for non-correlating source blur while using an uncorrelated noise model, such as that of Feng or Yu (Feng *et al* 2006, Yu *et al* 2000). This method would permit use of a non-linear objective function, permit use of Poisson (or other) noise models, eliminate bias imparted by the log transform, and potentially be more computationally efficient since the statistical weightings are independent. Alternately, one might adopt a non-staged reconstruction process where the entire forward model including system blurs is integrated into a nonlinear objective function. Such an approach would be attractive since this eliminates any parameter tuning associated with a deblurring step. These alternate models and reconstruction algorithms are the subject of ongoing work. While we have focused on 3D imaging in this paper, the 2D projection noise modeling of the processed measurements might also be applied to 2D restoration with the potential to improve projection radiography.

In addition to algorithm development, a major goal of future work will be more accurate modeling of system blurs, especially the higher order (high-frequency) properties of the source blur. The current blur model makes an assumption of shift-invariance. Both the shift-variant nature of the apparent focal spot size and shape, as well as the depth-dependent nature of source blur are the subject of ongoing studies. Specifically, it is relatively straightforward to generalize the implementation of the proposed work to handle a varying focal spot blur. For example, it is well known that the apparent focal spot size shrinks for positions on the detector that make larger angles with the piercing ray on the anode side of the tube; and the apparent focal spot size increases for increasing angles on the cathode side. Such variations can be incorporated into the ($\mathbf{B}_S$) term of the current model. Shift-variance due to depth-dependent source blur is potentially more complicated and computationally intense; and may require a modified forward projector or source models that use a collection of point sources.

The bench-data reconstructions shown here suggest applicability to current FP-CBCT systems. However, in addition to different blur properties, different systems and scans will have varying degrees of scatter, patient motion, and gantry jitter. Similarly, in this paper, we have focused on an extremity imaging example. We would expect other body sites to be potentially more challenging due to increased attenuation and likely increased scatter fractions. Our conjecture is that increased attenuation will predominantly increase noise, while increased scatter will likely reduce contrast but have relatively small effects on high-spatial resolution properties (since scatter effects are largely at low spatial frequencies). These properties will have to be investigated in more detail in future work to determine how they affect reconstruction image quality in real systems, how scatter and other artifact corrections interact with the proposed algorithm, and whether current correction schemes are sufficiently accurate for high resolution reconstructions.

In summary, we have presented a reconstruction method that breaks from the traditional assumption of spatially independent measurement noise. This is important since noise correlation due to flat-panel detectors is significant, and accurate noise models are a key element of MBIR methods. We have demonstrated improvements in the resolution-variance tradeoff, opening the opportunity for higher spatial resolution in flat-panel-based CBCT systems, including high-resolution extremities and breast imaging. We have also conducted preliminary investigations on the system designs that would benefit most from the proposed reconstruction method. This analysis is potentially important for future FP-CBCT system design since the proposed reconstruction method provides an alternate (software-based) route to achieving high spatial resolution. That is, the proposed methodology may permit alternate hardware designs (e.g. the ability to use larger focal spots with higher power limits) while still achieving the desired spatial resolution. Thus, this work has the potential to both extend the clinical performance of existing FP-CBCT systems and improve the tradeoffs and design choices for future clinical systems.

## 5 ACKNOWLEDGEMENTS

This work supported in part by NIH grants R21EB014964, T32EB010021, R01EB018896, and an academic-industry partnership with Varian Medical Systems (Palo Alto, CA). The authors would like to thank Sungwon Yoon and Kevin Holt for their many insights into this work. We would also like to thank Thomas Reigel for his assistance in collecting bench data.

## 6 APPENDIX

### 6.1 Covariance transformation

We show how the preprocessing steps transform the covariance matrix of the data. We start with the mean ($\bar{y}$) and covariance ($\mathbf{K}_Y$) of the measurement data. The deblurring and normalization steps are represented by the linear operators $\mathbf{C}'$ and $\mathbf{G}^{-1}$, respectively, resulting in deblurred normalized data ($y_{dbn}$) with mean and covariance given in (A1) and (A2) respectively.

$$\bar{y}_{dbn} = \mathbf{G}^{-1}\mathbf{C}'\bar{y} \tag{A1}$$

$$\mathbf{K}_{Y_{dbn}} = \mathbf{G}^{-1}\mathbf{C}'\mathbf{K}_Y[\mathbf{C}']^T[\mathbf{G}^{-1}]^T \tag{A2}$$

To calculate the effects of the log transform on the covariance matrix, we estimate $\log(x)$ as a linear function using a Taylor series expansion about $x_0$.

$$\log(x) \approx \log(x_0) + (x - x_0)/x_0 \tag{A3}$$

For each element of $y_{dbn}$ we expand about its current value

$$\log(y_i) \approx \log(y_{dbn,i}) + \frac{y_i - y_{dbn,i}}{y_{dbn,i}} \tag{A4}$$

$$\log(y) \approx \log(y_{dbn}) + \mathbf{D}\{y_{dbn}^{-1}\}y - 1 \tag{A5}$$

Addition of constants does not affect the covariance matrix, so the alterations to the covariance simply involve two multiplications by the diagonal matrix in the middle term in (A5). The covariance of the line integral estimates is given in (A6).

$$\mathbf{K}_L = \mathbf{D}\{y_{dbn}^{-1}\}\mathbf{G}^{-1}\mathbf{C}'\mathbf{K}_Y[\mathbf{C}']^T[\mathbf{G}^{-1}]^T\mathbf{D}\{y_{dbn}^{-1}\} \tag{A6}$$

Recognizing that **G** is diagonal, and the fact that

$$y_{dbn} = \mathbf{G}^{-1}\mathbf{C}'y \tag{A7}$$
$$\mathbf{D}\{y_{dbn}\} = \mathbf{G}^{-1}\mathbf{D}\{\mathbf{C}'y\} \tag{A8}$$
$$\mathbf{D}\{y_{dbn}^{-1}\} = \mathbf{G}\mathbf{D}\{\frac{1}{\mathbf{C}'y}\} \tag{A9}$$

Equation (A6) can be reduced to

$$\mathbf{K}_L = \mathbf{D}\{\frac{1}{\mathbf{C}'y}\}\mathbf{C}'\mathbf{K}_Y[\mathbf{C}']^T\mathbf{D}\{\frac{1}{\mathbf{C}'y}\} \tag{A10}$$